\documentclass[showpacs,prb,twocolumn]{revtex4}

\usepackage[english]{babel}
\usepackage{amssymb,amsmath}
\usepackage{color}
\usepackage{epsfig}
\usepackage{graphicx} 
\usepackage{ulem}

\begin{document}

\title{Many-body localized molecular orbital approach to molecular transport}

\author{Dmitry\,A.~Ryndyk\footnote{New affiliation: Institute for Materials Science, Dresden University of Technology, D-01069 Dresden, Germany}, Andrea~Donarini, Milena Grifoni, and Klaus~Richter}

\affiliation{Institute for Theoretical Physics, University of Regensburg,
D-93040 Regensburg, Germany
}

\begin{abstract}
An ab initio based theoretical approach to describe nonequilibrium many-body effects in molecular transport is developed. We introduce a basis of localized molecular orbitals and formulate the many-body model in this basis. In particular, the Hubbard-Anderson Hamiltonian is derived for single-molecule junctions with intermediate coupling to the leads. As an example we consider a benzenedithiol junction with gold electrodes. An effective few-level model is obtained, from which spectral and transport properties are computed and analyzed. Electron-electron interaction crucially affects transport and induces multiscale Coulomb blockade at low biases. At large bias, transport through asymmetrically coupled molecular edge states results in the occurrence of ``anomalous'' conductance features, i.e., of peaks with unexpectedly large/small height or even not located at the expected resonance energies.  
\end{abstract}

\pacs{\vspace{-0.1cm} 73.63.-b, 85.65.+h}

\maketitle

\section{Introduction}

Experimental and theoretical investigations of electron transport through single molecules are in the focus of the rapidly growing field of molecular electronics~\cite{Cuniberti05book,Cuevas10book,Song11advmat}. Electron-electron interaction plays an important role, controlling the position of resonant levels and leading to phenomena such as Coulomb blockade and Kondo effects. Depending on the ratio between the energy scales associated with an effective charging energy and coupling to the leads, molecular junctions can be classified in three groups. In the case of very small coupling, the molecular orbitals are only weakly disturbed, strong charge quantization and Coulomb blockade take place and transport is mainly determined by sequential tunneling~\cite{Park02nature,Kubatkin03nature,Osorio07advmat}. In the opposite case of large coupling, the electronic molecular orbitals are hybridized with  states in the leads, charge quantization is suppressed, transport is mainly coherent and the conductance is of the order of the conductance quantum $G_0=e^2/h$~\cite{Smit02nature,Tal08prl,Tal09prb}. Finally, in the intermediate situation, the effective electronic spectrum of a molecule is determined essentially by the hybridization, and the interplay between charge quantization and coherent transport may be important~\cite{Reed97science,Loertscher07prl,Kim11nanolett}. Let us consider as a commonly used benchmark example a gold-benzenedithiol-gold (Au-BDT-Au) molecular junction with the central molecule $S$-$C_6H_4$-$S$. The experiments~\cite{Reed97science,Loertscher07prl} show that, while it is difficult or impossible to observe the typical Coulomb blockade features in this type of molecular junctions, there are signatures of correlated transport, namely a conductance gap at small voltages and a complex shape of the conductance peaks. One can conclude that transport in the case of intermediate coupling can be correlated and partially incoherent. 

In parallel with the experimental investigations, a number of theoretical methods were suggested to describe the structure and electronic properties of molecular junctions. In the case of coherent transport ab initio DFT+NGF methods~\cite{Taylor01prb1,Damle01prb,Xue03prb1,Frauenheim02jpcm,Brandbyge02prb,Pecchia04repprogphys,Ke04prb}, combining density functional theory (DFT) and nonequilibrium Green function (NGF) techniques~\cite{Kadanoff62book,Keldysh64,Langreth76inbook,Rammer86RMP,Haug96book}, have become standard~\cite{Cuniberti05book,Cuevas10book}. However, the use of DFT, which is a powerful tool to deal with ground state properties, may become questionable when applied to transport and  nonequilibrium situations, especially when inelastic and interaction effects are essential. Indeed, the use of the solutions of the Kohn-Sham equation as physical quasiparticle states can not be rigorously established. Besides, the DFT+NGF is a mean-field theory and lacks to describe many-body effects in systems with strong electron interactions. Thus, being based on the basis of an atomistic modeling of molecular junctions, the rigorous extension of the DFT approach to describe transport phenomena remains a challenge.

\begin{figure}[b]
\centering
\includegraphics[width=0.5\textwidth]{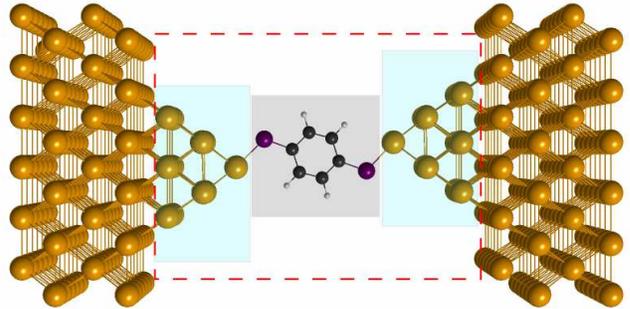} 
\caption{(Color online) Schematic picture of a Au-BDT-Au molecular junction. The dashed box defines the extended molecule. It comprises the central molecular region and parts of the leads, also marked with boxes.}
\label{bdt_316}
\end{figure}

On the other hand, model-based approaches are particularly important to understand the physics of correlated transport. The models are solved usually within two main approaches: the NGF technique in the case of strong coupling to the leads, and the quantum master equation (QME)~\cite{Blum96book,Breuer02book} in the weak-coupling limit. The quantum master equation is usually formulated in the basis of many-body eigenstates of the molecule. It gives a fairly complete description of sequential tunneling, the main features of Coulomb blockade and even can capture Kondo physics for temperatures of the order of or larger than the Kondo temperature~\cite{Koenig96prb}. Finally, for some nonperturbative effects covering the whole range of temperatures, e.g. the Kondo effect in the crossover regime, more sophisticated treatments are necessary, e.g. numerical renormalization group approaches~\cite{Costi94jpcm,Andergassen10nanotech}, other numerical methods~\cite{Segal10prb,Pletyukhov12prl} or Keldysh field theories~\cite{Smirnov1203.4360}. 

The challenge for molecular transport theory is to combine an ab initio approach, required to take into account a realistic geometry and capable to provide the electronic structure of molecular junctions, with a many-body quantum transport technique, which is necessary to incorporate the correlation effects. In this direction a number of different ab initio based approaches were suggested~\cite{Hettler03prl,Delaney04prl,Ferretti05prl,Mirjani11prb,Bergfield11nanolett,Frederiksen07prb,Thygesen07jcp,Thygesen08prb,Korytar10jpcm,Korytar11jpcm,Greuling11prb,Karolak11prl,Troester12prb,Strange11prb,Rangel11prb}. It should be noted that many-body calculations usually require sophisticated analytical and numerical methods and can be very time consuming. Hence the methods of transport theory can not usually be applied directly to large realistic systems; instead a combined approach is preferable, where an effective model takes into account only the states predominantly participating in transport. Progress in this direction was achieved recently in applications to Coulomb blockade phenomena~\cite{Delaney04prl}, correlated transport through atomic wires~\cite{Ferretti05prl,Mirjani11prb}, many-body interference~\cite{Bergfield11nanolett} and Kondo effect~\cite{Korytar10jpcm,Korytar11jpcm,Greuling11prb,Karolak11prl,Troester12prb}.  Besides, Coulomb blockade phenomena in benzene and benzenedithiol junctions were also considered on the basis of semiempirical atomistic models~\cite{Begemann08prb,Bergfield09prb,Sobczyk12prb}. However, a systematic ab initio based many-body theory of molecular transport is still lacking. 

Our aim is to further develop such a theory. The main problem to be solved on the way from an atomistic model to transport calculations is that a huge number of microscopic (single-atom) degrees of freedom must be reduced to an effective few-electron (or few many-body state) interacting model, as a prerequisite of successive many-body transport methods. The main building blocks of our approach are an effective electron model (an orthonormal basis of localized molecular orbitals and a many-body Hamiltonian in this basis) extracted from atomistic calculations, and a nonequilibrium quantum transport method, based on nonequilibrium Green functions or on the quantum master equation. 

In the case of strong or intermediate coupling to the leads the electronic molecular levels should be considered together with the levels in the leads. A corresponding generic atomistic model is shown in Fig.\,\ref{bdt_316}. A so-called extended molecule (inside the dashed box) is placed between equilibrium electrodes. The extended molecule consists of the central region (roughly the molecule) and two leads (the regions of electrodes near the molecule). The size and boundaries of the central region are actually not fixed and should be determined in a way to include all relevant electronic states, as we will see below. 

After having defined the appropriate size of the extended molecule, we proceed as follows:

i) We perform Hartree-Fock or DFT ab initio calculations within the extended molecule and obtain the molecular orbitals, which are orthogonal and can serve as a basis for a many-body Hamiltonian. 

ii) The central region (e.g. an organic molecule) and the metallic leads have quite different physical properties, and the approximations required to describe interactions are also different. Thus, it is convenient to transform the basis of molecular orbitals obtained in (i) into a basis of localized molecular orbitals (LMOs), which can be spatially separated into a basis for the central region and a basis in the lead ends. Besides, the advantage of LMOs is that the Coulomb interaction in this basis can be simplified to the Hubbard form. This procedure enables us to obtain a many-body Hubbard-Anderson Hamiltonian in the central region with ab initio calculated parameters, including the coupling to the leads. The rest of the leads can be treated within some mean-field approximation. 

iii) Using the Hubbard-Anderson Hamiltonian, many-body methods (nonequilibrium Green functions and the master equation in this paper) can be effectively applied to analyze spectral and transport properties of molecular junctions.

The paper is organized as follows. The ab initio basis of localized molecular orbitals is introduced in Sec.\,\ref{sec_lmo}. The electron-electron interaction and Hubbard approximation are discussed in Sec.\,\ref{sec_hubbard}. The parameters of the effective Hubbard-Anderson Hamiltonian for the subspace of electronic states relevant for transport through the central region are derived in Sec.\,\ref{sec_ham}. In Sec.\,\ref{sec_ngf} we use nonequilibrium Green functions to analyze the coupling to the electrodes and spectral properties. The many-body spectrum of the central region is discussed in Sec.\,\ref{sec:mbs} and the transport results are shown in Sec.\,\ref{sec:transport}. We give a short conclusion in Sec.\,\ref{sec:conclusion}.

\section{Localized molecular orbitals}
\label{sec_lmo}

The first stage of our approach includes ab initio calculations of the optimized geometry and the relevant basis of electronic states. For the preliminary geometry optimization and molecular dynamics of whole structures we apply the DFT code Siesta~\cite{Siesta}. For the extended molecule only (without the full electrodes) the full-electron quantum chemistry code Firefly (former PC GAMESS)~\cite{Firefly} is used. The final geometry optimization is performed using a hybrid DFT method, usually B3LYP. Then, the calculation of the molecular orbitals (MOs) is performed inside the extended molecule (including both the central  region and leads). Our test calculations show that a simple local density approximation (LDA) gives reasonable results for the considered systems. At this stage we obtain the MOs $\psi^{MO}_i({\bf r})$ with associated energies $E_i$. MOs have the advantage of being normalized and orthogonal. However, the canonical MOs of the extended molecule can not serve as a good basis for systems with interactions, because both electron-electron and electron-vibron interactions require physically different approximations in different, metal or molecular, parts of the junction. It is hence more convenient to use {\em localized} molecular orbitals (LMOs). The interactions between localized orbitals have a simple and transparent form and the appropriate approximations can be found. Besides, the number of required, relevant basis states is smaller and better controlled for LMOs than for MOs.     

\begin{figure}[t]
\centering
\includegraphics[width=0.4\textwidth]{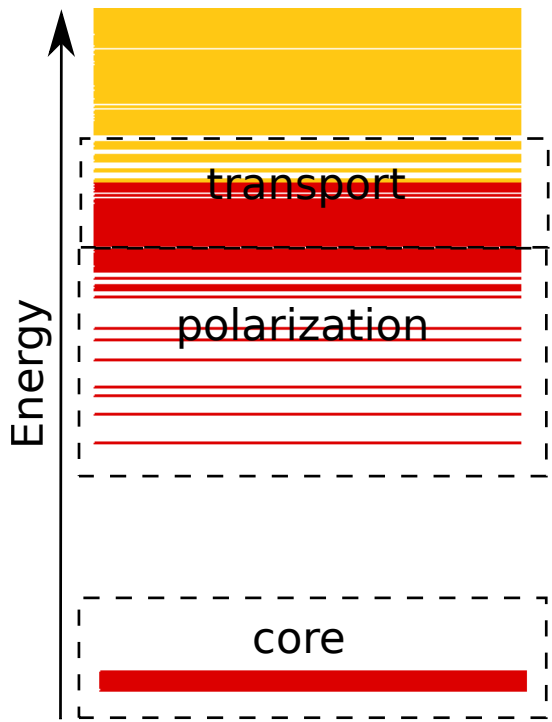}
\caption{(Color online) The energy spectrum of the molecular orbitals of the extended molecule shown in Fig.\,\ref{bdt_316}. The occupied levels are marked red, the empty levels yellow.}
\label{MOs}
\end{figure}

To proceed, we divide all MOs into four groups (Fig.\,\ref{MOs}). Most relevant for transport are the {\it transport} levels near the Fermi energy of the electrodes. These levels are selected for the localization procedure and include both occupied and valence MOs in an appropriate energy range around the Fermi energy. This energy range should be larger than the energy scales of the external bias voltage and temperature. The other criterion is that the obtained LMOs should be localized strongly enough inside the central region and in the leads: only in this case it is possible to separate the system into parts and use different approximations for the central region and leads separately. Alternatively, the partial localization of only relevant MOs (for example only $\pi$-type orbitals) is possible. In any case, these transport electron states play the main role. All other {\em polarization} states, which are further away from the Fermi energy or do not participate in transport for some other reasons, can still affect the interaction between transport electrons and, in particular, screen the Coulomb interaction. The polarization MOs can be localized in the same way as the transport orbitals. Finally, the core orbitals at low energies can be included into the effective (pseudo)potential and empty orbitals at high energies are neglected. 

\begin{figure}
\centering
\includegraphics[width=0.48\textwidth]{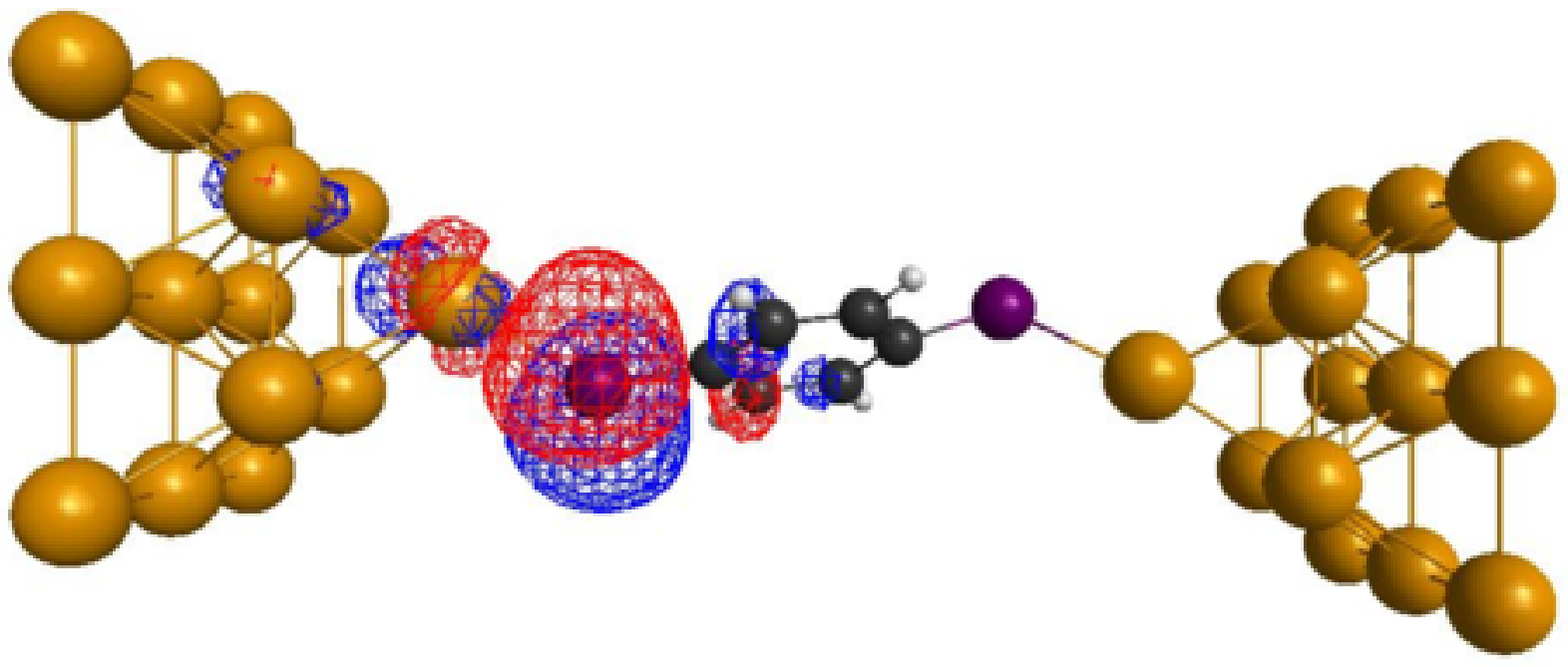} \\[-0.1cm]
\includegraphics[width=0.48\textwidth]{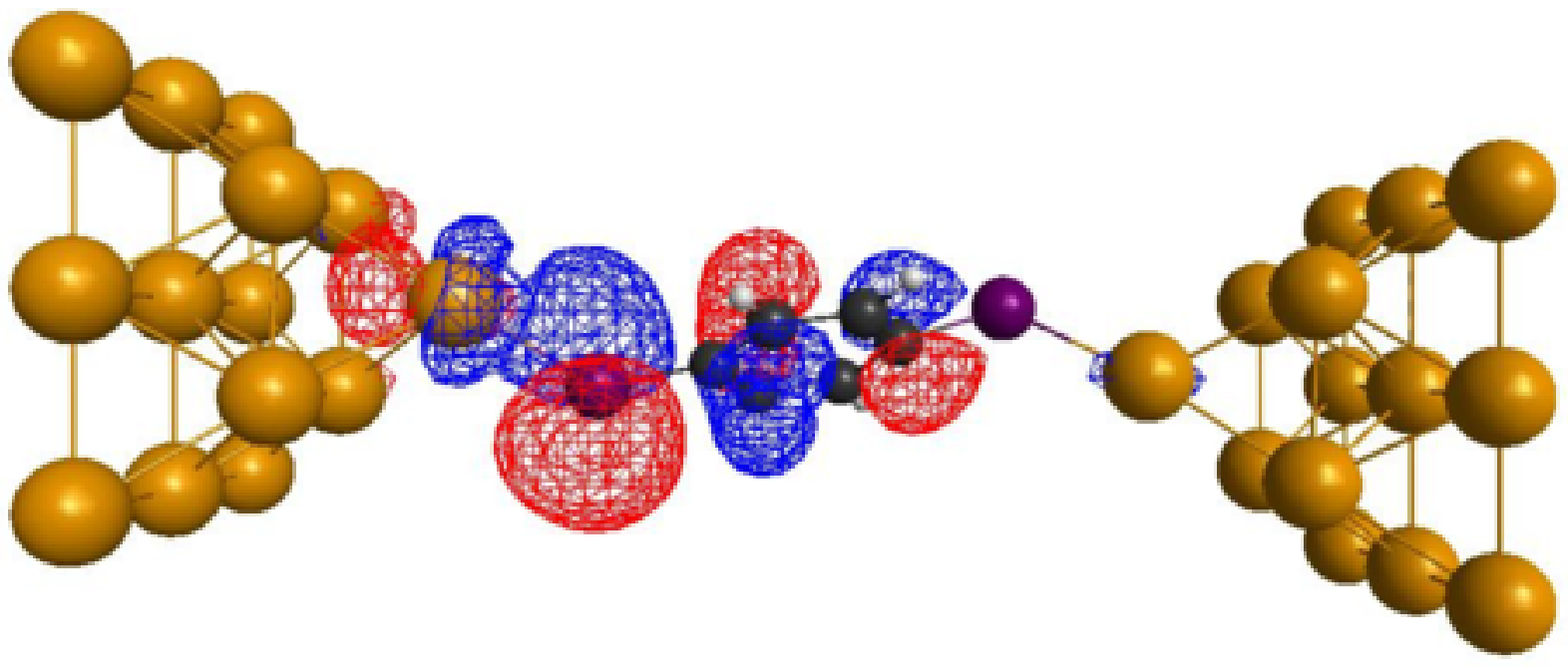} \\[-0.1cm]
\includegraphics[width=0.48\textwidth]{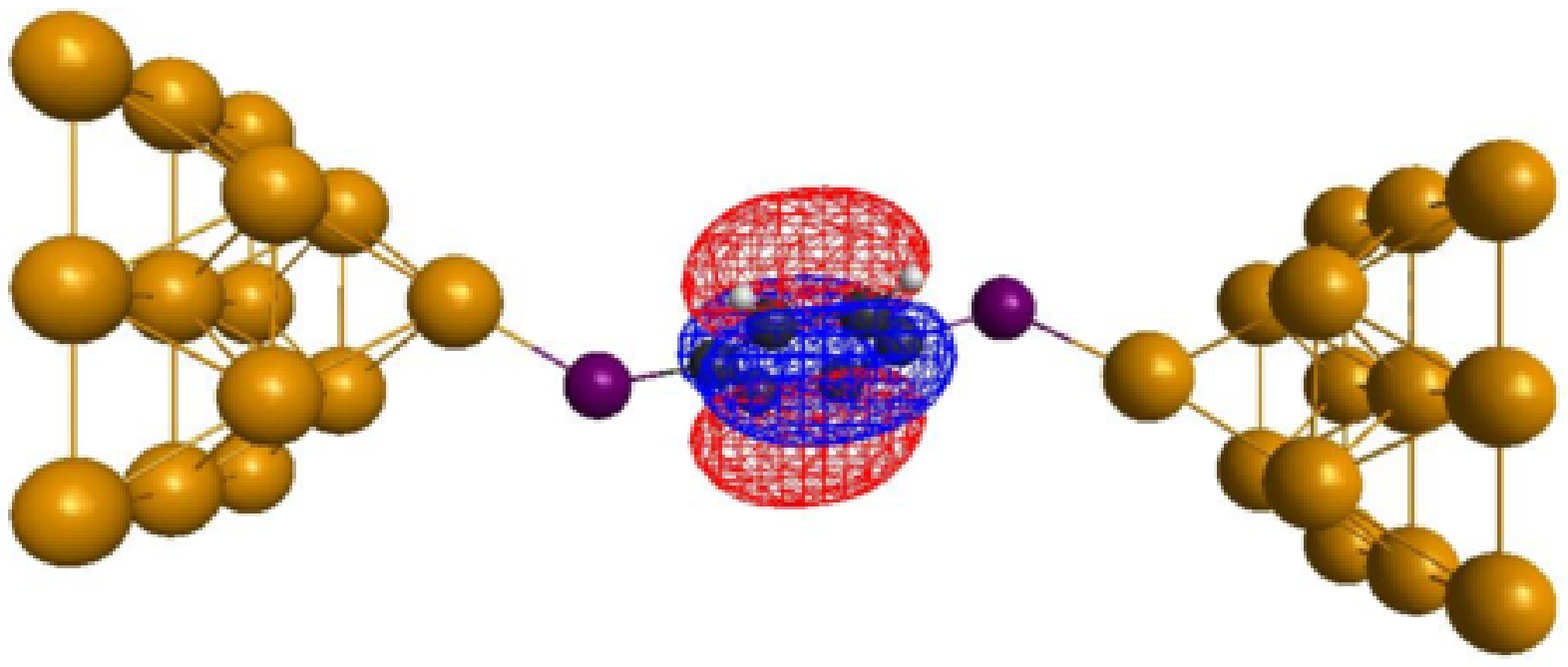} \\[-0.1cm]
\includegraphics[width=0.48\textwidth]{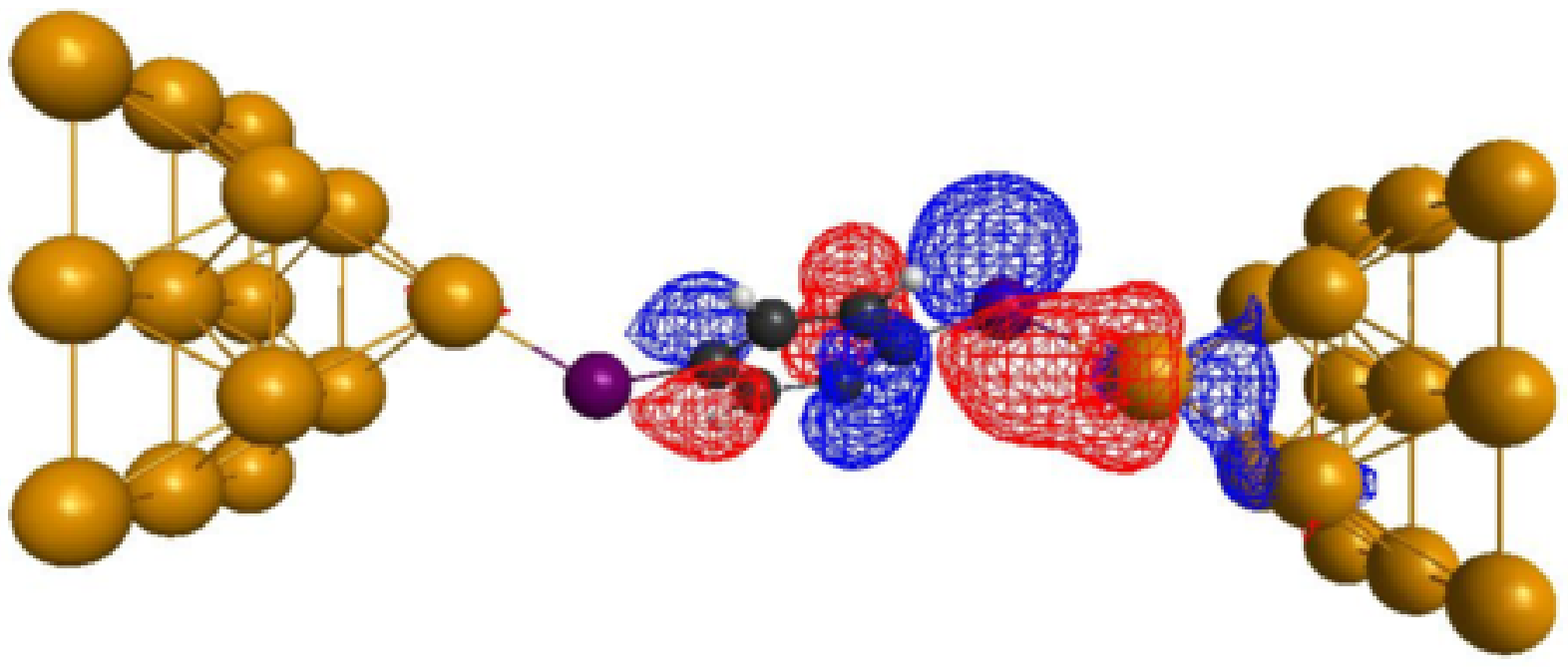} \\[-0.1cm]
\includegraphics[width=0.48\textwidth]{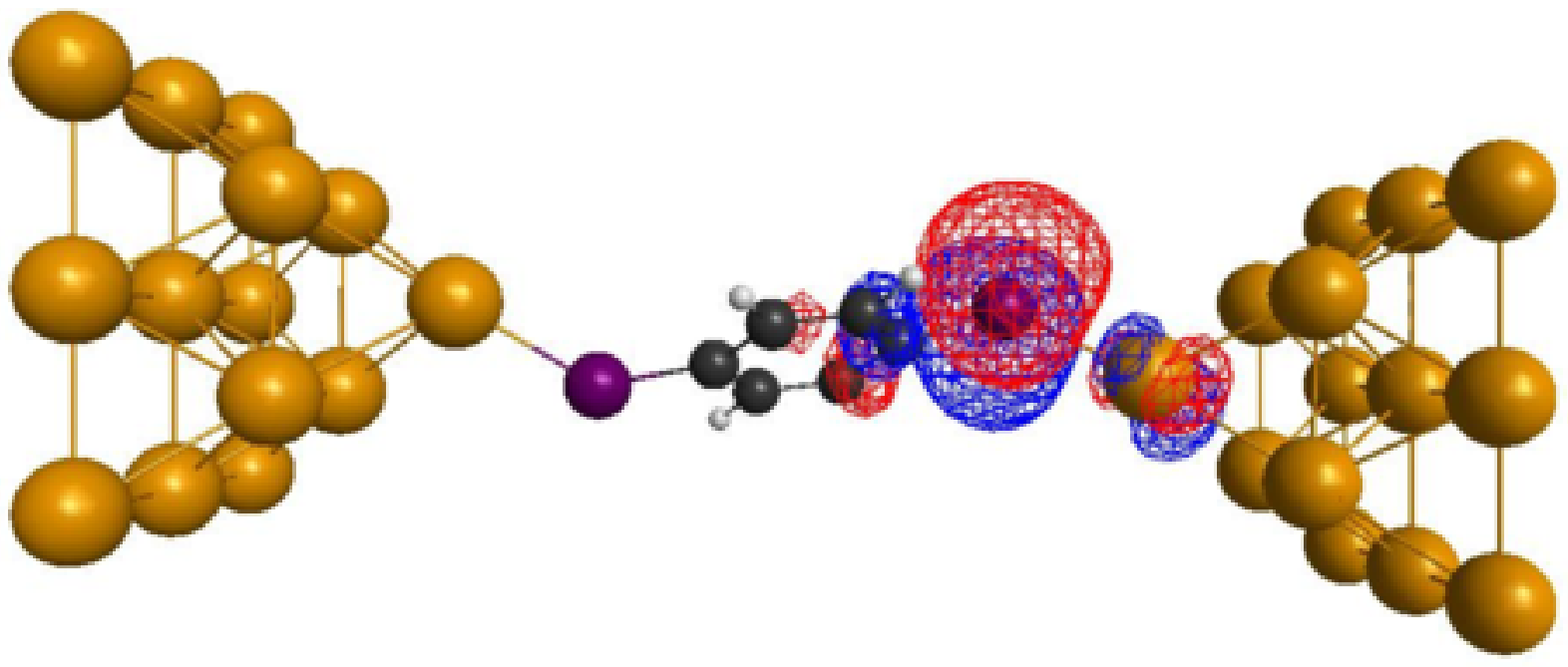} 
\caption{(Color online) Localized molecular orbitals in the central region of an Au-BDT-Au molecular junction.}
\label{bdt_lmos}
\end{figure}

\begin{figure} 
\includegraphics[width=0.5\textwidth]{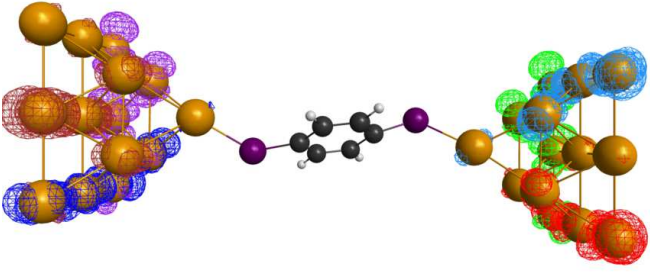}
\caption{(Color online) Localized molecular orbitals in the leads (several out of many are shown by different colors).}
\label{bdt_lmos_leads}
\end{figure}

For benzenedithiol, considered in this paper, the transport window was chosen to be about 4 eV, and contains about 60 MOs, mainly due to the large density of states in the metal leads. The parameters of the obtained effective model are rather robust against the exact choice of this number.  

The LMOs are obtained from the MOs by the unitary transformation
  \begin{equation}\label{mo2lmo}
    \psi^{LMO}_{\bar\alpha}({\bf r})=\sum_i s_{\bar\alpha i}\psi^{MO}_i({\bf r}).
  \end{equation}
The indices with bars $\bar\alpha,\bar\beta...$ denote the states without the spin degree of freedom. We apply the Foster-Boys localization method~\cite{Foster60rmp}, which minimizes the spatial extent of the orbitals and maximizes the distance between orbital centers. Thus, we obtain maximally localized orbitals. Out of the about 60 orbitals only 5 are localized in the central region. While the initial MOs spread across the extended molecule, the LMOs are spatially localized in the central region (Fig.\,\ref{bdt_lmos}) or in the leads (Fig.\,\ref{bdt_lmos_leads}).

Due to the unitary transformation (\ref{mo2lmo}) the LMOs are still orthogonal and normalized, but the expression
\begin{equation}\label{E_LMO}
  \epsilon^{LMO}_{\bar\alpha\bar\beta}=\sum_i s^{-1}_{\bar\alpha i}E_is_{i\bar\beta}
\end{equation}
is no longer diagonal. Moreover, the related Hamiltonian ${\hat H}^{LMO}$ takes the form
\begin{equation}\label{H_EM}
{\hat H}^{LMO}=\left(\begin{array}{ccc}{\hat H}_L & {\hat V}_{LC} & 0 \\
{\hat V}^\dag_{LC} & {\hat H}^{LMO}_C & {\hat V}^\dag_{RC} \\
0 & {\hat V}_{RC} & {\hat H}_R
\end{array}\right),
\end{equation}
where ${\hat H}_L$, ${\hat H}^{LMO}_C$, and ${\hat H}_R$ are the Hamiltonians of the left lead, the central region, and the right lead separately. The direct coupling between left and right leads can be neglected in most cases, because the LMOs of different leads are only very weakly overlapping.  

\section{Coulomb interaction and Hubbard approximation}
\label{sec_hubbard}

Having the LMOs at hand we can calculate the Coulomb matrix elements. The electron-electron interactions are described by the Hamiltonian
\begin{equation}\label{CB-H1}
  \hat H_{ee}=  \frac{1}{2}\sum_{\alpha\beta\gamma\delta}V_{\alpha\beta,\gamma\delta}
  d^\dag_\alpha d^\dag_\beta d_\delta d_\gamma ,
\end{equation}
where $\alpha=\{\bar\alpha,\sigma_\alpha\}$ and $\sigma_\alpha$ denotes the spin. The matrix elements for the scalar Coulomb interaction $ V(|{\bf r-r'}|)$ are defined as
\begin{align}\label{CB-H4}
&  V_{\alpha\beta,\gamma\delta}=V_{\bar\alpha\bar\beta,\bar\gamma\bar\delta}\delta_{\sigma_\alpha\sigma_\gamma}\delta_{\sigma_\beta\sigma_\delta}, \\
&  V_{\bar\alpha\bar\beta,\bar\gamma\bar\delta}=
  \int d{\bf r}\int d{\bf r'}\psi^*_{\bar\alpha}({\bf r})\psi^*_{\bar\beta}({\bf r'})
  V(|{\bf r-r'}|)\psi_{\bar\gamma}({\bf r})\psi_{\bar\delta}({\bf r'}),
\end{align}
where $\delta_{\sigma_\alpha\sigma_\beta}$ is the Kronecker symbol. For the systems with localized wave functions $\psi_\alpha({\bf r})$, where the overlap between two different states is weak, the main matrix elements are those with $\bar\alpha=\bar\gamma$ and $\bar\beta=\bar\delta$. We checked that, indeed, the overlap of 3 or 4 different orbitals can be neglected in comparison with the overlap of only 2 orbitals.  In these cases it suffices to replace $\hat H_{ee}$ by the Hubbard Hamiltonian
\begin{equation}\label{CB-AH}
  \hat H_{H}=\frac{1}{2}\sum_{\alpha\neq\beta}U_{\alpha\beta}\hat n_\alpha \hat n_\beta,
\end{equation}
describing only density-density interactions with \mbox{$\hat n_\alpha=d^+_\alpha d_\alpha$} and the Hubbard parameters defined as
\begin{equation}\label{LMO_U}
  U_{\alpha\beta}=
  \int d{\bf r}\int d{\bf r'}|\psi^{LMO}_{\bar\alpha}({\bf r})|^2|\psi^{LMO}_{\bar\beta}({\bf r'})|^2
  V(|{\bf r-r'}|).
\end{equation}
As a further advantage of LMOs, the local nature of electron correlation is better described in the localized basis.

The interaction $V(|{\bf r-r'}|)$ is the bare Coulomb interaction \mbox{$V(r)=1/r$}, if we include all electrons into the localization procedure. Actually we restrict the effective Hamiltonian to transport electronic states, on which we performed the Hubbard approximation. The remaining polarization electrons are included only through a screened Coulomb interaction. The screening can be described by the effective interaction potential in RPA (or GW)  approximation, which is, however, energy dependent. To keep the simplicity of the Hubbard approximation, we use a screened Coulomb interaction \mbox{$V(r)=1/(\varepsilon r)$} with $\varepsilon\approx 1.5$. This approximation gives reasonable values of $U_{\alpha\beta}$ for the benchmark $\pi$-orbital model of benzene C$_6$H$_6$ which we compared with the optimized semi-empirical PPP model~\cite{PPP}. In this way all coefficients are derived, but further semi-empirical corrections could be included for better agreement with experiments.   

\section{Many-body model}
\label{sec_ham}

\begin{figure}[b]
\centering
\includegraphics[width=0.4\textwidth]{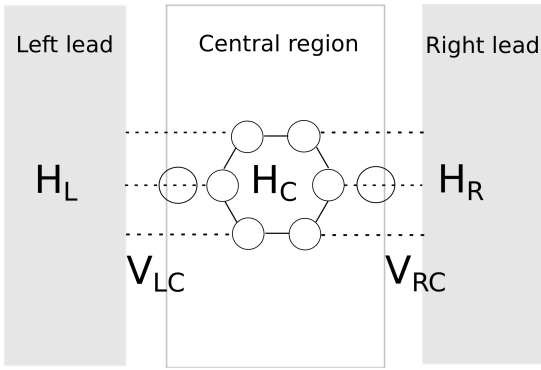}
\caption{The model of the extended molecule.}
\label{SMJ_mod_1}
\end{figure}

The next important step is the derivation of the entire effective Hamiltonian in the basis of the LMOs for the central region. As we already explained, we divide the extended molecule into the central part (actually the molecule in our particular case) and the leads  (Fig.\,\ref{SMJ_mod_1}). The full Hamiltonian is the sum of the noninteracting molecular Hamiltonian $\hat H^0_C$, the electron-electron interaction Hamiltonian $\hat H_{ee}$, the Hamiltonians of the ends of the leads $\hat H_{R(L)}$, and the tunneling Hamiltonian $\hat H_T$ describing the molecule-to-lead coupling:
\begin{equation}
 \hat H=\hat H^0_C+\hat H_{ee}+\hat H_L+\hat H_R+\hat H_T.
\end{equation}
In our case the central Hamiltonian $\hat H^0_C+\hat H_{ee}$ is replaced by the Hubbard cluster Hamiltonian:
\begin{equation}\label{AHM}
\hat{H}_C=\hat H^0_C+\hat H_{H}= \sum_{\alpha\beta}\tilde\epsilon_{\alpha\beta}\hat{d}^\dag_\alpha\hat{d}_\beta+\frac{1}{2}\sum_{\alpha\neq\beta}U_{\alpha\beta}\hat{n}_\alpha\hat{n}_\beta,
\end{equation}
where $\tilde\epsilon_{\alpha\beta}=\epsilon_{\alpha\beta}+e\varphi_\alpha\delta_{\alpha\beta}$ are the bare energies of the LMOs, including the shifts due to an external voltage.

The noninteracting molecular Hamiltonian $\hat H^0_C$ is obtained from the LMO Hamiltonian ${\hat H}^{LMO}_C$, Eq.\,(\ref{H_EM}).  The zero-voltage energies $\epsilon_{\alpha\beta}=\epsilon_{\bar\alpha\bar\beta}\delta_{\sigma_\alpha\sigma_\beta}$ are calculated from the HF or DFT MO energies $E_i$, from which the contribution of the Hartree terms due to Hubbard interactions should be substracted:
  \begin{equation}
    \epsilon_{\bar\alpha\bar\beta}=\sum_i s^{-1}_{\bar\alpha i}E_is_{i\bar\beta}-\Delta\epsilon_{\bar\alpha\bar\alpha}\delta_{\bar\alpha\bar\beta},
  \end{equation}
where 
  \begin{align}\nonumber
  \Delta\epsilon_{\bar\alpha\bar\alpha} & =U_{\bar\alpha\bar\alpha}n^{0}_{\bar\alpha}+\sum_{\bar\gamma\ne\bar\alpha}2U_{\bar\alpha\bar\gamma}n^0_{\bar\gamma}  \\ \label{DeltaEps}
  & =\sum_{i}^{N_{occ}}\sum_{\bar\gamma}|s_{i\bar\gamma}|^2\left[U_{\bar\alpha\bar\alpha}\delta_{\bar\alpha\bar\gamma}+2U_{\bar\alpha\bar\gamma}(1-\delta_{\bar\alpha\bar\gamma})\right].
  \end{align}
In this expression $n^{0}_{\bar\alpha}$ denote the populations of the corresponding LMOs in the ab initio calculation. The sum is taken over all occupied molecular orbitals.

The coupling to the leads is described by the tunneling Hamiltonian 
\begin{equation}\label{H_tun}
 \hat H_T=\sum_{s=L,R}\sum_{k\sigma,\alpha}\left(V^*_{sk\sigma,\alpha}
 c^{\dag}_{sk\sigma}d_\alpha+V_{sk\sigma,\alpha}d_\alpha^{\dag}c_{sk\sigma}\right),
\end{equation}
and the Hamiltonians of the left and right leads are
\begin{equation}\label{H_leads}
 \hat H_{s=L(R)}=\sum_{k\sigma}\tilde\epsilon_{sk\sigma}
 c^{\dag}_{sk\sigma}c_{sk\sigma},
\end{equation}
where $k$ is the index of a state, and $\sigma$ is the spin. Note that in our case the states in the leads are not plane waves, but are represented by LMOs, calculated simultaneously with the LMOs in the active region. The leads are considered at the mean-field (DFT) level. The equilibrium electrodes, which can have different electrochemical potentials, determine the boundary conditions for the leads.

For the 5-level model (Fig.\,\ref{bdt_lmos}), which represents actually 10 levels with spin, we obtain the following parameters (for spin up or down, all numbers are in $eV$):
\begin{equation}
\label{L5_H}  
\{\epsilon_{\bar\alpha\bar\beta}\}=\left(\begin{array}{cccccc} 
  -17.80 &   0.06 &    0.00 &    0.03 &    0.06 \\
    0.06 & -19.81 &   -0.04 &   -0.54 &   -0.02 \\
    0.00 &  -0.04 &  -18.79 &    0.04 &   -0.06 \\
    0.03 &  -0.54 &    0.04 &  -18.20 &    0.00 \\
    0.06 &  -0.02 &   -0.06 &    0.00 &  -16.25
\end{array}\right).
\end{equation}
The matrix of the Hubbard parameters calculated from expression (\ref{LMO_U}) reads 
\begin{equation}
\label{L5_U} 
\{U_{\bar\alpha\bar\beta}\}=\left(\begin{array}{cccccc} 
   4.08 &   2.72 &    2.31 &    1.36 &    1.36 \\
   2.72 &   4.32 &    2.43 &    2.34 &    1.36 \\
   2.31 &   2.43 &    3.84 &    1.59 &    1.59 \\
   1.36 &   2.34 &    1.59 &    4.00 &    2.72 \\
   1.36 &   1.36 &    1.59 &    2.72 &    4.08
\end{array}\right).  
\end{equation}
The diagonal entries in this matrix correspond to the interaction of two electrons in the same LMO state but with different spin. The off-diagonal terms denote coupling beween two different LMOs spin of the electronic state. Of course, these expressions should be transformed into the 10-level basis before performing further calculations.

At finite bias voltage $V$ (defined by the left and right electrical potentials, \mbox{$V=\varphi_L-\varphi_R$}) the energies are shifted. In linear approximation these shifts are described by $\eta_\alpha$: $\tilde\epsilon_{\alpha\beta}=\epsilon_{\alpha\beta}+e\varphi_\alpha\delta_{\alpha_\beta}$, $\varphi_\alpha=\varphi_R+\eta_\alpha(\varphi_L-\varphi_R)$,
where the parameters $0 < \eta_\alpha < 1$ characterize the symmetry of the voltage drop across the junction, $\eta_\alpha=0.5$ stands for the symmetric case. Note that the energies and other parameters can also be {\em ab initio} calculated at finite voltage, but that is very time-consuming.

The coupling matrix elements $V_{sk\sigma,\alpha}$ in Eq.\,(\ref{H_tun}) are obtained directly from the localization procedure. Indeed, the Hamiltonian of an extended molecule takes the form Eq.\,(\ref{H_EM}), and the off-diagonal terms describe the coupling to the leads. The number of states in the leads is many times larger than in the central region. Thus, to leading approximation, we can average over the lead level distributions and couplings (so-called wide-band limit). In this approximation the level-width function
\begin{equation}\label{LWF}
  \Gamma_{s,\alpha\beta}(\epsilon) =2\pi\sum_{k\sigma}V_{sk\sigma,\beta}V^*_{sk\sigma,\alpha}\delta(\epsilon-\tilde\epsilon_{sk\sigma})
\end{equation}
is energy independent.

We now return to our 5-level model. The couplings of the first two states (localized at the left sulfur atom) and the last two states (localized at the right sulfur atom) are characterized by the level-width functions $\Gamma_{L11}=0.16\,eV$, $\Gamma_{L22}=0.21\,eV$, $\Gamma_{R44}=0.28\,eV$, $\Gamma_{R55}=0.1\,eV$. All other couplings are small and do not play an essential role. Thus, all parameters of the model Hamiltonian Eqs.\,(\ref{AHM})-(\ref{H_leads}) are well defined and we can proceed with transport calculations. In view of the experiments~[\onlinecite{Reed97science,Loertscher07prl}], we will perform calculations below room temperature, $k_BT\leq 0.025\,eV$, implying $\Gamma>k_BT$.  

As we discussed in the introduction, transport at finite voltage can be described in the framework of nonequilibrium Green function or quantum master equation approaches implying numerical methods. For benzene-based junctions several methods were used, including coherent DFT based, the master equation approach in the sequential tunneling limit~\cite{Hettler03prl,Begemann08prb}, sophisticated approximations in the framework of the nonequilibrium Green function method~\cite{Thygesen08prb,Bergfield09prb,Strange11prb,Rangel11prb}, and other methods~\cite{Delaney04prl}. In this paper we use both NGF and QME methods, trying to attack the problem from both sides. It should be noted, however, that for our case, $k_BT<\Gamma\ll U$, both NGF and a QME with second order rates can only give a qualitative description of the transport problem. Very recently, a QME approach for a single-level junction able to describe the regime $\Gamma\sim k_BT$ has been proposed~\cite{Kern1209.4995}. An extension of this method to a multilevel system will be subject of future investigations. 

\section{Nonequilibrium Green function approach to spectral properties}
\label{sec_ngf}

We follow the formulation pioneered by Meir, Wingreen and Jauho ~\cite{Meir92prl,Jauho94prb,Jauho06jpcs}. The lesser (retarded, advanced) Green function matrix ${\hat G}^{<(R,A)}\equiv G_{\alpha\beta}^{<(R,A)}$ of a nonequilibrium molecule can be found from the Dyson-Keldysh equations in the integral form
\begin{eqnarray}\label{GR_Int}
& {\hat G}^R(\epsilon)={\hat G}^R_0(\epsilon)
  +{\hat G}^R_0(\epsilon){\hat\Sigma}^R(\epsilon)
  {\hat G}^R(\epsilon), \\[0.3cm]
& {\hat G}^<(\epsilon)=
  {\hat G}^R(\epsilon){\hat\Sigma}^<(\epsilon){\hat G}^A(\epsilon),
\end{eqnarray}
or from the corresponding equations in the differential form
\begin{equation}\label{GR_Diff}
  (\epsilon-\tilde\epsilon_{\alpha\alpha})G^R_{\alpha\beta}-\sum_\gamma
\Sigma^{R}_{\alpha\gamma} G^R_{\gamma\beta}=\delta_{\alpha\beta},
\end{equation}
\begin{equation}\label{GK_Diff}\begin{array}{c}\displaystyle
  (\tilde\epsilon_{\beta\beta}-\tilde\epsilon_{\alpha\alpha})G^<_{\alpha\beta}- \sum_\gamma
 \left(\Sigma^{R}_{\alpha\gamma}G^<_{\gamma\beta}+\Sigma^{<}_{\alpha\gamma}G^A_{\gamma\beta}
  - \right. \\
\left.-G^{R}_{\alpha\gamma}\Sigma^<_{\gamma\beta}
-G^{<}_{\alpha\gamma}\Sigma^A_{\gamma\beta}   \right) =0. \end{array}
\end{equation}
Here
\begin{equation}
{\hat\Sigma}^{R,A,<}= {\hat\Sigma}^{R,A,<(T)}_{L}+{\hat\Sigma}^{R,A,<(T)}_{R}+{\hat\Sigma}^{R,A,<(I)}
\end{equation}
is the total self-energy of the molecule composed of the interaction self-energy ${{\hat\Sigma}}^{R,A,<(I)}$ and the tunneling (coupling to the left (L) and right (R) lead) self-energies
\begin{eqnarray}\label{SigmaR_{HF}}
&\displaystyle {\hat\Sigma}_{s=L,R}^{R,<(T)}={\hat V}^\dag_{sC}{\hat G}^{R,<}_s {\hat V}_{sC}, \\
&\displaystyle {\Sigma}_{s\alpha\beta}^{R,<(T)}=\sum_{k\sigma}\left\{V^*_{sk\sigma,\alpha}
 {G}^{R,<}_{sk\sigma}V_{sk\sigma,\beta}\right\},
\end{eqnarray}
where ${G}^{R,A,<}_{sk\sigma}$ is the Green function of the leads.

The retarded tunneling  self-energy ${\hat\Sigma}_s^{R(T)}$ can be represented as
\begin{equation}\label{SigmaRT}
{\hat\Sigma}^{R(T)}_{s}(\epsilon)={\hat\Lambda}_s(\epsilon-e\varphi_s)-\frac{i}{2}
{\hat\Gamma}_{s}(\epsilon-e\varphi_s),
\end{equation}
where ${\hat\Lambda}_s$ is the real part of the self-energy, which usually can be included
in the single-particle Hamiltonian $\hat H^0_C$, and ${\hat\Gamma}_s$ describes level
broadening due to coupling to the leads. In the case of noninteracting leads with continuous energy spectra, the level-width function is determined by the expression (\ref{LWF}).

For the corresponding lesser function of the noninteracting leads one finds
\begin{equation}
{\hat\Sigma}^{<(T)}_s(\epsilon)=i{\hat\Gamma}_s(\epsilon-e\varphi_s)
f^0_s(\epsilon-e\varphi_s),
\end{equation}
where 
\begin{equation}
f_s^0(\epsilon)=\frac{1}{\exp\left((\epsilon-\mu_s)/T\right)+1}
\end{equation}
is the equilibrium Fermi distribution function with chemical potential $\mu_s$ ($k_B=1$).

The expression for the interaction self-energy can not be obtained exactly. In the nonequilibrium Hartree-Fock approximation one has \cite{Ryndyk09inbook}
\begin{align}
& \Sigma^{R(I)}_{\alpha\beta}= \left(\sum_\gamma U_{\alpha\gamma} \langle\hat n_\gamma\rangle\right)\delta_{\alpha\beta}, \\
& \Sigma^{<(I)}_{\alpha\beta}=0.
\end{align}
We do not consider more sophisticated cases here.

The current from the left ($s=L$) or right ($s=R$) lead into the central system is described by the  expression (for spin-unpolarized leads)
\begin{equation}\label{J}\begin{array}{c}\displaystyle
 J_{s=L,R}=\frac{\mathrm{i}e}{\hbar}\int\frac{d\epsilon}{2\pi}{\rm Tr}\left\{
 {\hat\Gamma}_s(\epsilon-e\varphi_s)\left({\hat G}^<(\epsilon)+ \right.\right.\\[0.5cm]
 \displaystyle \left.\left.
 f^0_s(\epsilon-e\varphi_s)
 \left[{\hat G}^R(\epsilon)-{\hat G}^A(\epsilon)\right]\right)\right\},
 \end{array}
\end{equation}

Applying the NGF technique to our case, we should take into account that our system initially consists not of three, but of five parts: the large electrodes, quantum leads and the central region (Fig.\,\ref{bdt_316}). Accordingly, the full Hamiltonian has the form   
\begin{equation}
{\hat H}=\left(\begin{array}{ccccc}
{\hat H}^{el}_L & {\hat V}^{el}_L & 0 & 0 & 0 \\
{\hat V}^{el\dag}_L & {{\hat H}_L} & {{\hat V}_{LC}} & {0} & 0 \\
0 & {{\hat V}^\dag_{LC}} & {{\hat H}_C} & {{\hat V}^\dag_{RC}} & 0 \\
0 & {0} & {{\hat V}_{RC}} & {{\hat H}_R} & {\hat V}^{el\dag}_R \\
0 & 0 & 0 & {\hat V}^{el}_R & {\hat H}^{el}_R 
\end{array}\right),
\end{equation}
where the central part is the same as before, Eq.\,(\ref{H_EM}), and the additional terms describe the electrodes and the coupling between the electrodes and the leads.  

\begin{figure}[t]
\includegraphics[width=0.5\textwidth]{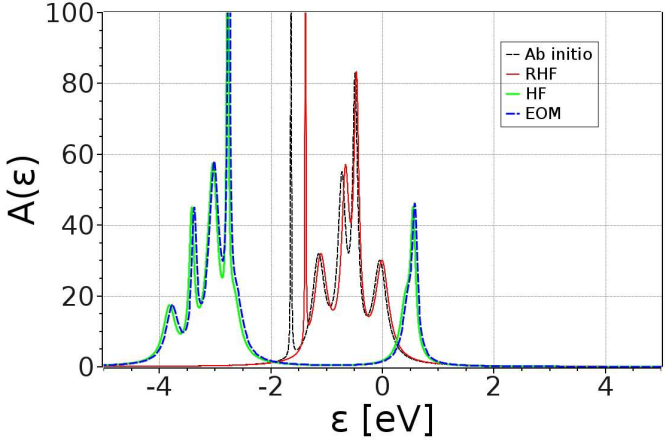}
\caption{(Color online) The spectral function within the different approximations: initial DFT (black dashed), restricted HF (red), unrestricted HF (green) and equation-of-motion (dashed blue).}
\label{A}
\end{figure}

The solution of Eq.\,(\ref{GR_Int}) for the central part is in this case
\begin{equation}
  {{\hat G}^R_{C}(\epsilon)}=\frac{1}{(\epsilon+i\eta){\hat I}-{{\hat H}^0_C}{-{\hat\Sigma}^{R(T)}_L-{\hat\Sigma}^{R(T)}_R
  -{\hat\Sigma}^{R(I)}}},
\end{equation}
with the lead self-energies $\hat\Sigma^{R(T)}_s$ (see Eq.\,(\ref{SigmaR_{HF}})) and level-width functions defined as
\begin{eqnarray}\label{Sig1}
\displaystyle  {\hat\Sigma}^{R(T)}_L= {\hat V}^\dag_{LC}{\hat G}^R_L {\hat V}_{LC},\ \ \ \  {\hat\Gamma}_L=-2{\rm Im}{\hat\Sigma}^{R(T)}_L, \\
\label{Sig2}
\displaystyle  {\hat\Sigma}^{R(T)}_R= {\hat V}^\dag_{RC}{\hat G}^R_R {\hat V}_{RC},\ \ \ \  {\hat\Gamma}_R=-2{\rm Im}{\hat\Sigma}^{R(T)}_R.
\end{eqnarray}
The lead Green functions for noninteracting leads (or for leads described by the effective mean-field Hamiltonians ${\hat H}_s$)  are defined correspondingly by the electrode self-energies
\begin{eqnarray}
&\displaystyle  {\hat G}^{R}_{L}(\epsilon)=\frac{1}{(\epsilon+i\eta){\hat I}-{{\hat H}_L}-{\hat\Sigma}^{el}_L}, \\
&\displaystyle  {\hat G}^{R}_{R}(\epsilon)=\frac{1}{(\epsilon+i\eta){\hat I}-{{\hat H}_R}-{\hat\Sigma}^{el}_R}.
\end{eqnarray}
The calculation of the electrode self-energies is done by standard methods~\cite{Cuniberti05book,Cuevas10book}. The lesser Green functions are calculated in the same way. We assume additionally that the distribution function in the leads is the same as in the corresponding electrodes.

The equations are solved self-consistently within four approximations: initial DFT  with the mean-field energies Eq.\,(\ref{E_LMO}), the restricted HF approximation (RHF) with $n_{\bar\alpha\uparrow}=n_{\bar\alpha\downarrow}$, the unrestricted HF approximation (\ref{SigmaR_{HF}}) and, finally, the equation of motion method (EOM)~\cite{Song07prb,Ryndyk09inbook}. 

First we analyze the equilibrium (zero bias) spectral function of the central region,
\begin{equation}
A(\epsilon)=-2\sum_\alpha{\rm Im}G^R_{C\alpha\alpha}. 
\end{equation}
The first thing one can see (Fig.\,\ref{A}) is that the RHF approximation gives a spectral function similar to the ab initio (DFT) one. This is not surprising as the ab initio calculation is also RHF and we simply extracted the Hartree contribution when calculating the energies $\epsilon_{\bar\alpha\bar\beta}$ in Eq.\,(\ref{DeltaEps}). The other important point is that the results obtained in HF and EOM approximations are distinctly different and a gap is opened at the Fermi surface. The analysis of the populations $n_\alpha=\langle\hat n_\alpha\rangle$ of the single-particle states shows that two empty states are located at the left and one at the right side of the molecule (second and fifth states in Fig.\,\ref{bdt_lmos}). These two empty states have the same spin (in the HF approximation the ground state is spin polarized and degenerate, but quantum fluctuations can switch between different spin orientations), indicating that the true ground state, with quantum fluctuations taken into account, can be spin singlet or triplet. We discuss this point in the next section.  

\section{The many-body spectrum: \newline exact diagonalization and the ground state properties}
\label{sec:mbs}

To gain insight into the many-body energy spectrum of the central system we perform an exact diagonalization of Eq.~(\ref{AHM}) obtaining the set of many-body eigenstates $|\lambda\rangle$. Calculating the tunneling matrix elements we obtain from Eqs. (\ref{AHM}) and (\ref{H_tun}) the Hamiltonian
\begin{align}
\hat H_C+\hat H_T=\sum_\lambda & E_\lambda |\lambda\rangle\langle\lambda| \\ +\sum_{s=L,R;k\sigma;\lambda\lambda'}
 &  \left[T_{sk\sigma,\lambda'\lambda}|\lambda'\rangle\langle\lambda|c_{sk\sigma}+T^*_{sk\sigma,\lambda'\lambda}|\lambda\rangle\langle\lambda'|c^\dag_{sk\sigma}
\right], \nonumber \\
& T_{sk\sigma,\lambda'\lambda}=\sum_\alpha V_{sk\sigma,\alpha}\langle\lambda'|{d}^\dag_\alpha|\lambda\rangle.
\end{align}

First, we analyze the many-body spectrum. With 5 LMOs we get 1024 many-body states in the Fock space. The lowest 8-particle states consist of a series of alternating singlets and triplets (see Table\,\ref{tab:levels}). In particular, the ground state is a singlet, practically degenerate with a triplet ($E_{8'} - E_{8_g} \approx 10^{-4} eV$). It follows, at a distance of roughly $0.3 eV$ a second pair of singlet-triplet quasi degenerate states. The 9-particle states are all doublets with a relatively regular distance of the order of $0.5 eV$. Finally, it is important to keep in mind that the energies of the lowest four 8-particle levels lie all below the one of the 9-particle ground state. The 7-particle states have much higher energies.
\begin{table}[h]
\begin{tabular}{|c|c|c|}
 \hline
 Level& Energy [eV] & Spin [$\hbar$] \\ 
 \hline
 $8_g$& -91.1849 & 0 \\
 $8'$ & -91.1848 & 1 \\
 $8''$ & -90.8653 & 0 \\
 $8'''$ & -90.8648 & 1 \\
 $9_g$ & -90.7866 & 1/2 \\
 $9'$ & -90.3693 & 1/2 \\
  $9''$ & -90.0891 & 1/2 \\
  \hline
\end{tabular}
\caption{The eigenenergies and the associated spins of the lowest 8 and 9 particle levels of an Au-BDT-Au molecular junction.}
\label{tab:levels}
\end{table}

Note that in the sequential-tunneling master equation method the exact many-body states can be partially occupied at finite temperatures, but not at zero temperature, and the level broadening is not taken into account. This can give some noticeable difference in the position of the transport resonances compared to the NGF calculations, where the levels can be partially occupied even at low temperatures, and where the real part of the HF self-energy, Eq.\,(\ref{SigmaR_{HF}}), describes the energy shift of the single-particle levels. 

The population probabilities $P_\lambda$ are found from the master equation
\begin{align}\label{me1}
  \frac{dP_\lambda}{dt}=\sum_{\lambda'}&\left(\Gamma^{\lambda\lambda'}P_{\lambda'}-\Gamma^{\lambda'\lambda}P_{\lambda}\right),
\end{align}
where the tunneling rates are, in second order in the tunneling Hamiltonian, 
\begin{align}\nonumber
  \Gamma^{\lambda\lambda'}=\sum_{s=L,R;\sigma}&\left[\gamma^{s\sigma}_{\lambda\lambda'}f^0_\sigma(E_{\lambda}-E_{\lambda'}-e\varphi_s) \right. \\ 
  &\left.+\gamma^{s\sigma}_{\lambda'\lambda}\left(1-f^0_\sigma(E_{\lambda'}-E_{\lambda}-e\varphi_s)\right)\right],
  \label{me2}
\end{align}
with
\begin{align} \nonumber
&  \gamma^{s\sigma}_{\lambda\lambda'}=\frac{2\pi}{\hbar}\sum_{k}T_{sk\sigma,\lambda\lambda'}T^*_{sk\sigma,\lambda\lambda'}
   \delta(E_\lambda-E_{\lambda'}-\tilde\epsilon_{sk\sigma}) \\ \nonumber
&  =\frac{2\pi}{\hbar}\sum_{\alpha\beta k}V_{sk\sigma,\beta}\langle\lambda|\hat{d}^\dag_\beta|\lambda'\rangle
    V^*_{sk\sigma,\alpha}\langle\lambda'|\hat{d}_\alpha|\lambda\rangle\delta(E_\lambda-E_{\lambda'}-\tilde\epsilon_{sk\sigma}) \\ 
&  =\frac{2\pi}{\hbar}\sum_{\alpha\beta}\Gamma_{s\sigma,\alpha\beta}(E_\lambda-E_{\lambda'})\langle\lambda|\hat{d}^\dag_\beta|\lambda'\rangle
    \langle\lambda'|\hat{d}_\alpha|\lambda\rangle. 
    \label{me3}
\end{align}
This expression connects the tunneling rates to the level-width function; thus the $\Gamma_{s\sigma,\alpha\beta}(\epsilon)$ calculated by the NGF method can be used, see Eqs.\,(\ref{Sig1},\ref{Sig2}). In the wide-band limit one has $\hbar\gamma^{s\sigma}_{\lambda\lambda'}=2\pi\rho_0|T_{s\sigma,\lambda\lambda'}|^2$, where $\rho_0$ is the density of states.

To check that the simple (diagonal) form of the master equation can be used, we have analyzed the many-body spectrum of the considered system and came to the conclusion that no coherences are needed for the description of the transport since the degeneracies are not of orbital but of spin nature (e.g. triplets for the 8-particle and doublets for the 9-particle states). However, there cannot be mixing of states with different total spin since otherwise the mixing will depend on the choice of the direction of the quantization axis. The solution of the Eqs.\,(\ref{me1}-\ref{me3}) is straightforward and can be obtained by direct numerical integrations in stationary and time-dependent cases.

As we discussed before, in the equilibrium state at zero voltage there are 8 electrons distributed due to thermal smearing between the states $8_g$ and $8'$, see Tab.\,\ref{tab:levels}. An equilibrium occupation with 8 electrons is in agreement with the HF calculations of Fig.\,\ref{A}. In Table\,\ref{tab:n} the composition of the many-body states in terms of the five localized molecular orbitals of Fig.\,\ref{bdt_lmos} is quantified in terms of the average populations $n_\alpha=\langle\hat n_\alpha\rangle$ of the single particle states obtained from exact diagonalization of $H_C$ (see Eq.\,(\ref{AHM})). The composition of the states $8_g$ and $8'$ is similar to the HF average populations. As discussed in Sec.\,\ref{sec:transport} and shown in Fig.\,\ref{fig:Q_LMO}, the LMO occupations change at finite bias.

\begin{table}[h]
\begin{tabular}{|c|c|c|c|c|c|}
  \hline
   & $n_1$ & $n_2$ & $n_3$ & $n_4$ & $n_5$ \\
   \hline
   NGF  & 1.9725 & 1.0934 & 1.9989 & 1.9294 & 1.0220 \\
   \hline
  $8_g$ & 1.9801 & 1.1337 & 1.9973 & 1.8685 & 1.0204 \\
  $8'$ & 1.9798 & 1.1339 & 1.9972 & 1.8685 & 1.0207 \\
  $8''$ & 1.0390 & 1.8047 & 1.9997 & 1.1753 & 1.9815 \\
  $8'''$ & 1.0269 & 1.8159 & 1.9995 & 1.1764 & 1.9812 \\
  \hline
  $9_g$ & 1.9990 & 1.4657 & 1.9992 & 1.5365 & 1.9996 \\
  $9'$ & 1.9510 & 2.0000 & 1.9989 & 1.9999 & 1.0502 \\
  $9''$ & 1.0772 & 1.9917 & 1.9998 & 1.9780 & 1.9532 \\
  \hline
\end{tabular}
\caption{The average populations $n_\alpha$ of the single-particle states at zero bias voltage. Calculations within the NGF method are shown in the second line. They agree with the composition of the states $8_g$ and $8'$ obtained from exact diagonalization of $H_C$ (see Eq.\,(\ref{AHM})).}
\label{tab:n}
\end{table}


\section{Transport characteristics}
\label{sec:transport}

\begin{figure}[t]
\includegraphics[width=0.5\textwidth]{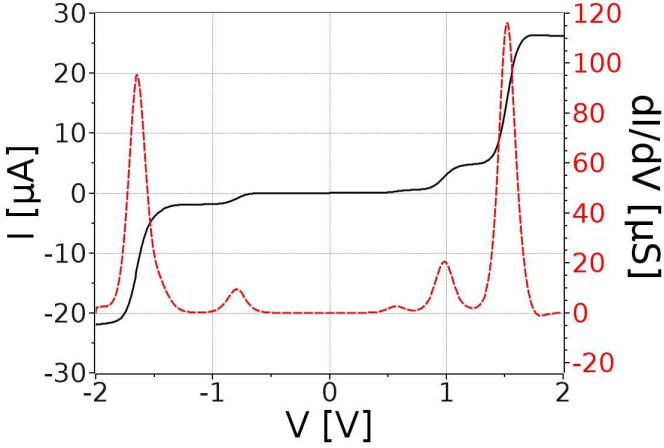}
\caption{(Color online) Current-voltage curve (black solid line) and differential conductance (red dashed line).}
\label{IV_QME}
\end{figure}

We are now able to calculate and interpret the current-voltage characteristics of the benzenedithiol junction. The current at finite voltage, which is given by 
%
\begin{align}\nonumber
  I_{s=L,R}=e&\sum_{\lambda\lambda';\sigma}\left[\gamma^{s\sigma}_{\lambda\lambda'}
  f^0_\sigma(E_{\lambda}-E_{\lambda'}-e\varphi_s)P_{\lambda'} \right.\\ 
  &\left.-\gamma^{s\sigma}_{\lambda\lambda'} \left(1-f^0_\sigma(E_{\lambda}-E_{\lambda'}-e\varphi_s)\right)P_{\lambda} \right],
\end{align}
is presented together with the differential conductance in Fig.\,\ref{IV_QME}. The curves are asymmetric with respect to a bias inversion because the junction geometry was chosen to be slightly asymmetric. As the main result we find a multi-scale Coulomb blockade. The large region of suppressed current is about 2 Volt wide. However, the current is completely blocked only in a much smaller region of bias voltage, as small steps in the current (peaks in the conductance) are present at lower biases.

\begin{figure}[t]\centering
\includegraphics[width=0.5\textwidth]{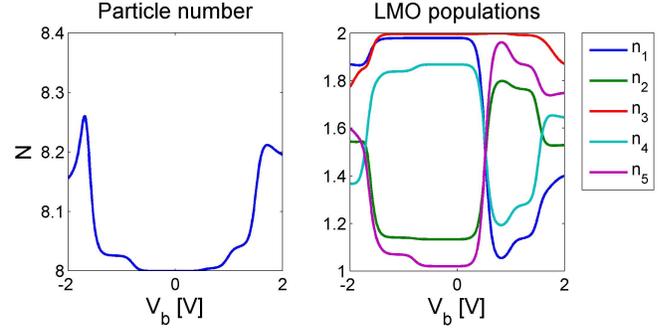}
\caption{Bias dependence of the average electron number (left) and average individual populations (right) on the molecular junction.}
\label{fig:Q_LMO}
\end{figure}

As a first step in the analysis of the current voltage characteristics we consider the average particle number in the central system presented in the left panel of Fig.\,\ref{fig:Q_LMO}. At low biases the average particle number is 8 corresponding to the neutral configuration of benzenedithiol. The many-body state with the minimal grand-canonical energy ($E_G = E -\mu N$) is in fact the 8-particles ground state (see Fig.\,\ref{fig:Dynamics}). When the bias drop is raised in the junction the average particle number takes values between 8 and 9 ensuring that the dominant transitions are negative ion resonances. 

A further insight into the dynamics is obtained by monitoring the average occupation of the different localized molecular orbitals shown in the right panel of Fig.\,\ref{fig:Q_LMO}. At low biases the symmetric central orbital (the third orbital from the top in Fig.\,\ref{bdt_lmos}) is completely occupied, $n_3 = 2$. Its occupation undergoes a sensible variation only at the voltages of the large current steps $V_b \approx \pm  1.5 V$. Large variations in the population of the asymmetric LMOs centered around the molecule-lead interface (orbitals 1, 2, 4  and 5 in Fig. 3) are instead associated to the small current steps present at lower voltages. Interestingly, at a bias of $V_b \approx 0.5 V$ the effective spatial symmetry of the system is recovered with the populations of the asymmetric states being all equal.

\begin{figure}[b]\centering
\includegraphics[width=0.4\textwidth]{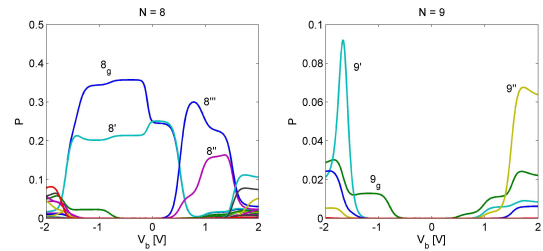}
\caption{The occupation of the manybody states}
\label{fig:MBS}
\end{figure}

\begin{figure*}[t]\centering
\includegraphics[width=\textwidth]{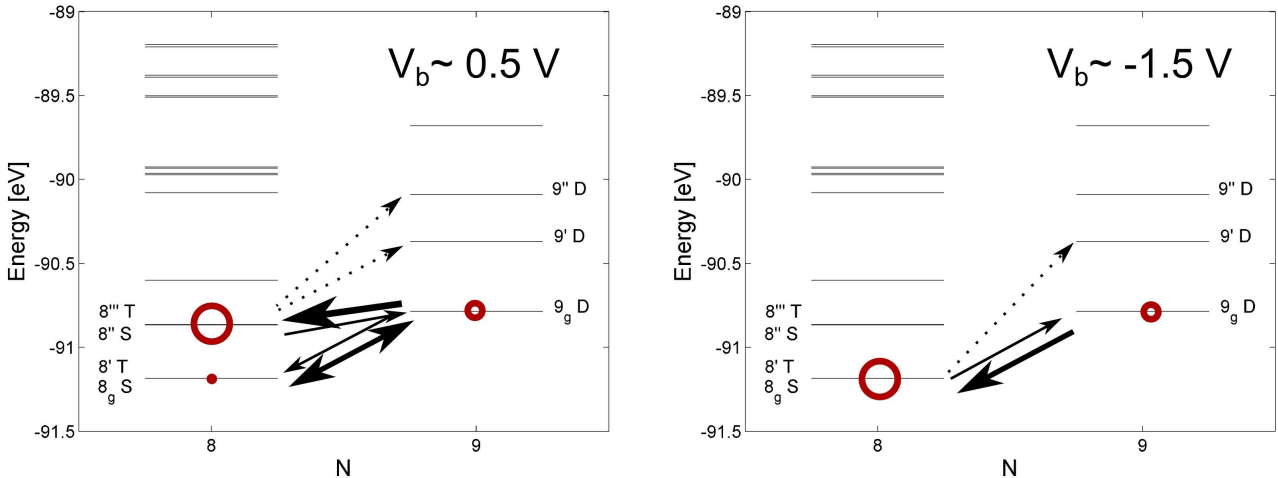}
\caption{The energies, tunneling rates and the associated populations of the most relevant states (see discussion in the text).}
\label{fig:Dynamics}
\end{figure*}

A deeper understanding of the dynamics of the system is obtained by the analysis of the occupation of the many-body states (Fig. \ref{fig:MBS}), their energies (Tab.\,\ref{tab:levels}), and the transition rates among them (Tab.\,\ref{tab:rates}, schematically represented in Fig.\,\ref{fig:Dynamics}). If the calculation of the current is performed taking into account hundreds of many-body states, the essential physics at the biases presented in Fig.\,\ref{IV_QME} is captured by considering the lowest four 8-particle levels (for a total of 8 states) and the lowest three 9-particle ones (6 states).

The tunnelling events from (to) the source or the drain connect these many-body states. The tunnelling rate $\Gamma^{\lambda\lambda'}$ is the product of a geometrical part ($\gamma^{s\sigma}_{\lambda\lambda'}$ of Eq.\,(\ref{me3})) and an energetic contribution encoding the energy conservation in the tunnelling event and the Pauli exclusion principle (see Eq.\,(\ref{me2})).
The energetic contribution ensures that the rate $\Gamma^{\lambda\lambda'}$ changes (and correspondingly the current through the system) every time the resonant condition $E_{\lambda} - E_{\lambda'} - e\phi_s = 0$ is fulfilled. With this argument it is already possible to assign a specific transition to most of the peaks in the conductance of Fig.\,\ref{IV_QME}. In particular the transitions $8_g, 8' \leftrightarrow 9'$ are associated with the peak at the most negative bias and $8_g \leftrightarrow 9_g$ to the second peak from the left. The first small peak at positive bias is anomalous and we will return to it later. We only note that its position depends on the temperature and that it moves to the $8_g \leftrightarrow 9_g$ resonance at low temperatures. The rightmost conductance peaks are instead associated to the transitions $8'', 8''' \leftrightarrow 9'$ and $8'',8''' \leftrightarrow 9''$, respectively.

The approximate symmetries of molecular geometry are very important since they introduce selection rules which distinguish between transitions which are energetically equally allowed. In Tab.\,\ref{tab:rates} we report the transition rates $\gamma^{s\sigma}$ between the different many-body states. Here the values are given in $eV$ and the spin $\sigma$ is chosen to fulfill spin conservation in the tunnelling event. In the case of a doublet to triplet transition the value of the rate reported is the one involving the triplet state with maximum projection along the quantization axis. Except for the transition $8' \leftrightarrow 9''$ all transitions show a very pronounced left-right asymmetry. It is much easier for example for an electron to tunnel in (or out) of the molecule from (to) the left instead of the right lead when this tunnelling event involves the many-body eigenstates $8'$ and $9'$. This asymmetry is essential to explain the dynamics of the system at low biases and can be understood in terms of the spatial distribution of the many-body eigenstates. 

\begin{table}[t]
\begin{tabular}{|c|c|c|c|}
  \hline
  $\gamma^{L\sigma}$ & $9_g$ & $9'$ & $9''$ \\
  \hline
  $8_g$ & 0.0017 & 0.1025 & 0.0024 \\
  $8'$ & 0.0056 & 0.2039 & 0.004 \\
  $8''$ & 0.1033 & 0.0003 & 0.0074 \\
  $8'''$ & 0.1676 & 0.0013 & 0.0488 \\
  \hline
\end{tabular}
\\
\begin{tabular}{|c|c|c|c|}
  \hline
  $\gamma^{R\sigma}$ & $9_g$ & $9'$ & $9''$ \\
  \hline
  $8_g$ & 0.0442 & 0.0127 & 0 \\
  $8'$ & 0.0866 & 0.0278 & 0.0022 \\
  $8''$ & 0.0044 & 0.0056 & 0.1157 \\
  $8'''$ & 0.0013 & 0.011 & 0.2185 \\
  \hline
\end{tabular}
\caption{The transition rates $\gamma^{s\sigma}$ between the different manybody states.}
\label{tab:rates}
\end{table}

The left transition rate is larger than the right one when the transition from an 8 to a 9-particle state is associated to a larger variation of the density in the orbitals 1 or 2 than in the orbitals 4 or 5. Analogous arguments hold for the reverse situation.

Let us now return to the interpretation of the current voltage characteristics with the help of Fig.\,\ref{fig:Dynamics}. By convention, to a positive bias voltage $V_b$ corresponds a stationary particle current flowing from right to left while the electrical current flows from left to right. We concentrate first on the negative bias. From an accurate analysis of the definition of the tunnelling rates (Eq.\,(\ref{me2})) it is not difficult to prove that the first step in the current is due to the
resonant condition between the $8_g (8')$ and $9_g$ states at the left lead. Current flows since the system oscillates between the $8_g (8')$ and $9_g$ states by receiving an electron from the left lead and by releasing it to the right one. The asymmetry between the transition rates, $\gamma^{R\sigma}>\gamma^{L\sigma}$, ensures than even after the opening of the current channel the occupation of the $8_g$ (together with the almost degenerate $8'$) is still the largest one. In the right panel of Fig.\,\ref{fig:Dynamics} we schematically represent the tunnelling rates and the associated populations of the most relevant levels for a bias just above (in absolute value) the first negative bias conductance peak. Starting from this population distribution it is then natural to observe the next visible current step related to the transition $8_g (8') \leftrightarrow 9'$. Since this time the left tunnelling rate dominates, the population of the 8-particle states decreases substantially in favor of the 9-particle ones. Generally, a more uniform mixing of states with different particle number is associated with a larger fluctuation of the number of electrons in the central system and thus with a larger current.  

The dynamics at positive bias is more complex. In particular the first conductance peak occurs at a bias at which even the ground to ground state transition is not yet open. This anomalous behaviour is understandable when taking into account the large left-right asymmetry of the rates. As schematically represented in the left panel of Fig.\,\ref{fig:Dynamics}, even before the (right lead) resonance between the $8_g (8')$ and the $9_g$ state opens a conventional current channel, the states $8''$ and $8'''$ get strongly populated. The fundamental reason is the large probability to tunnel out of the system at the left lead through the transition $9_g \to $ $8'', 8'''$ which is also energetically favorable. Very soon the states $8''$ and $8'''$ become the new effective ground states for the system (see Fig.\,\ref{fig:MBS}). In this scheme it is thus not surprising that i) the first conductance peak is located at an "average" between the $8_g(8') \leftrightarrow 9_g$ resonance and the $8''(8''') \leftrightarrow 9_g$ one; ii) the next two conductance peaks at positive bias occur at the $8''(8''') \leftrightarrow 9'$ and $8''(8''') \leftrightarrow 9''$ resonant conditions.

\section{Conclusions}
\label{sec:conclusion}

In conclusion, we developed a many-body localized molecular orbital approach to transport through molecular junctions with the following protocol:

\begin{enumerate}

\item Geometry optimization using DFT and hybrid DFT (usually B3LYP based) methods.

\item Molecular vibrons can be calculated after the geometry optimization (not considered in this paper).

\item Molecular orbitals of the extended molecule are obtained. Localized molecular orbitals (LMOs) are constructed and form the basis for all subsequent calculations. 

\item A Hubbard interaction is introduced for the LMOs in the central region: only density-density Coulomb integrals are taken into account. 

\item Electron-vibron interaction can be included in the central region (not considered in this paper).

\item The leads are kept as effectively noninteracting (mean-field approximation). The interaction Hamiltonian between leads and central region yields the relevant tunneling couplings.  

\item A spectral analysis and transport calculations are performed on the basis of the ab initio based Hubbard-Anderson model.

\end{enumerate}

Using the benchmark example of a benzene-dithiol molecular junction, we performed the full line of calculations in the framework of this approach. We determined the geometry of the junction, calculated molecular orbitals and transformed them into localized molecular orbitals. Upon using an energy range of about 4 eV around the Fermi energy of gold, we obtained a basis of 5 LMOs with energies $\epsilon_{\alpha\beta}$. Then we calculated the Coulomb matrix elements $U_{\alpha\beta}$ for these orbitals and coupling matrix elements $V_{sk\sigma,\alpha}$ between the central region and the leads. Using the parameters $\epsilon_{\alpha\beta}$, $U_{\alpha\beta}$ and $V_{sk\sigma,\alpha}$, obtained from ab initio calculations, we calculated the spectral function in the framework of the nonequilibrium Green function approach (in the RHF, HF and NEOM approximations). Besides, the  model was transformed into the many-body eigenstate basis, and the quantum master equation (applied in the sequential tunneling limit) was used to calculate the current. It is shown that transport through asymmetrically-coupled molecular edge states results in suppressed peaks of the differential conductance at small voltage and unexpectedly large peaks at higher voltages. The origin of these anomalies could be explained upon analyzing the occupation probabilities of the many-body states as well as their compositions in terms of LMOs. In general, we could qualitatively understand the equilibrium state and main transport properties of the considered molecular junction with strong electron-electron interaction and intermediate coupling to the leads.    

Nevertheless, the further development of the theory is necessary with respect to both ab initio and quantum transport aspects. The results presented in this paper are only partially self-consistent because the parameters $\epsilon_{\alpha\beta}$, $U_{\alpha\beta}$ and $V_{sk\sigma,\alpha}$ are calculated at zero voltage, but used at all voltages. Actually, it is possible to extend the theory to include the recalculation of the parameters at finite voltage and the influence of the nonequilibrium charge in the central region on the leads. A related issue is the effect of the external field on the LMOs energies, which we treat using a simplified linear approximation. The Hubbard interaction plays the main role, but the corrections due to non density-density interactions and polarization of the molecule can be important as well. Finally, we expect that the method proposed in Ref.~\onlinecite{Kern1209.4995} could be of importance to treat the parameter regime $k_BT<\Gamma<U$ typical for  molecular junctions with intermediate coupling to the leads. 

\section*{Acknowledgments}

The {\em ab initio} calculations were done by  the quantum chemistry code Firefly~\cite{Firefly} and partially by the DFT code Siesta~\cite{Siesta}. The results were analyzed and the LMOs were visualized with the help of MacMolPlt~\cite{MacMolPlt} and Chemcraft. Many-body modeling and transport calculations were performed by our own codes. We thank Michael Hartung for his help with the local Linux cluster used for the numerical calculations. 

This work was funded by Deutsche Forschungsgemeinschaft within the Priority Program SPP 1243 and Collaborative Research Center SFB 689.


\end{document}